\documentclass[twocolumn,showpacs,preprintnumbers,amsmath,amssymb]{revtex4}
\usepackage{graphicx}
\usepackage{dcolumn}
\usepackage{bm}

\usepackage{amsmath}
\renewcommand{\vec}[1]{\mbox{\boldmath $#1$}}
\newcommand{\sgn}{\mbox{sgn}}

\begin{document}
\preprint{APS/123-QED}
\title{Statistical mechanics of lossy compression using multilayer perceptrons}
\author{Kazushi Mimura}
\affiliation{
Faculty of Information Sciences, Hiroshima City University, Hiroshima 731-3194, Japan}
\email{mimura@cs.hiroshima-cu.ac.jp}
\author{Masato Okada}
\affiliation{
Graduate School of Frontier Sciences, University of Tokyo, Chiba 277-5861, Japan \\
Brain Science Institute, RIKEN, Saitama 351-0198, Japan \\
PRESTO, Japan Sciences and Technology Agency, Chiba 277-8561, Japan 
}
\date{\today}
\begin{abstract}
\label{abst}
Statistical mechanics is applied to lossy compression using multilayer perceptrons for unbiased Boolean messages. 
We utilize a tree-like committee machine (committee tree) and tree-like parity machine (parity tree) whose transfer functions are monotonic. 
For compression using committee tree, a lower bound of achievable distortion becomes small as the number of hidden units $K$ increases. 
However, it cannot reach the Shannon bound even where $K\to\infty$. 
For a compression using a parity tree with $K \ge 2$ hidden units, the rate distortion function, which is known as the theoretical limit for compression, is derived where the code length becomes infinity. 
\end{abstract}
%
\maketitle


\section{introduction}

 Cross-disciplinary fields that combine information theory with statistical mechanics have developed rapidly in recent years and achievements in these have become the center of attention. 
The employment of methods derived from statistical mechanics has resulted in significant progress in providing solutions to several problems in information theory, 
including problems in error correction \cite{Sourlas1989,Kabashima2000,Nishimori1999,Montanari2000}, spreading codes \cite{Tanaka2001,Tanaka2005} and compression codes \cite{Murayama2003,Murayama2004,Hosaka2002,Hosaka2005}. 
Above all, data compression plays an important role as one of the base technologies in many aspects of information transmission. 
Data compression is generally classified into lossless compression and lossy compression \cite{Shannon1948,Shannon1959,Cover1991}. 
Lossless compression is aimed at reducing the size of message under the constraint of perfect retrieval. 
In lossy compression, on the other hand, the length of message can be reduced by allowing a certain amount of distortion. 
The theoretical framework  for lossy compression scheme is called rate distortion theory, which consists partly of Shannon's information theory \cite{Shannon1948,Shannon1959}. 
\par
Several lossy compression codes, whose schemes saturate the rate distortion function that represents an optimal performance, were discovered in the case where the code length becomes infinity. 
For instance, Low Density Generator Matrix (LDGM) code \cite{Murayama2003,Murayama2004} and using a nonmonotonic perceptron \cite{Hosaka2002,Hosaka2005,Hosaka2005b} were proposed. 
In these compression codes, a decoder is first defined to retrieve a reproduced message from a codeword. 
In the encoding problem, for a given source message, we must find a codeword that minimizes the distortion between the reproduced message and the source message. 
Therefore, fundamentally, the computational cost of compressing a message is of exponential order of a codeword length. 
It is important to understand properties of various lossy compression codes saturating the optimal performance for the development of more useful codes. 
\par
Since a multilayer network includes a nonmonotonic perceptron as a special case, 
we employ tree-like committee machine and parity machine as typical multilayer networks \cite{Barkai1990,Barkai1991,Barkai1992} to lossy compression and analytically evaluate their performance.

\section{lossy compression}

\begin{figure}[h]
  \vspace{10mm}
  \begin{center}
  \includegraphics[width=0.45\linewidth,keepaspectratio]{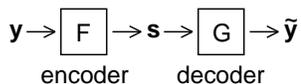}
  \end{center}
  \caption{Rate distortion encoder and decoder.}
  \label{fig:framework}
  \vspace{5mm}
\end{figure}

Let us start by defining the concepts of the rate distortion theory \cite{Cover1991}. 
Let $y$ be a discrete random variable with source alphabet ${\cal Y}$. 
We will assume that the alphabet is finite. 
An source message of $M$ random variables, $\vec{y}={}^t(y^1,\cdots,y^M)\in{\cal Y}^M$, is compressed into a shorter expression, 
where the operator ${}^t$ denotes the transpose. 
Here, the encoder describes the source sequence $\vec{y}\in{\cal Y}^M$ by a codeword $\vec{s}={\cal F}(\vec{y})\in{\cal S}^N$. 
The decoder represents $\vec{y}$ by a reproduced message $\hat{\vec{y}}={\cal G}(\vec{s})\in\hat{\cal Y}^M$, as illustrated in Fig. \ref{fig:framework}. 
Note that $M$ represents the length of a source sequence, while $N$ represents the length of a codeword. 
The code rate is defined by $R=N/M$ in this case. 
A distortion function is a mapping $d:{\cal Y}\times\hat{\cal Y}\to\mathbb{R}^+$ from the set of source alphabet-reproduction alphabet pair into the set of non-negative real numbers. 
In most cases, the reproduction alphabet $\hat{\cal Y}$ is the same as the source alphabet ${\cal Y}$. 
After this, we set ${\cal Y}=\hat{\cal Y}$. 
An example of common distortion functions is Hamming distortion given by 
\begin{equation}
d(y,\hat{y}) = \biggl\{
\begin{array}{ll}
0, & y=\hat{y}, \\
1, &y\ne\hat{y},
\end{array}
\end{equation}
which results in the probability of error distortion, since  $E[d(y,\hat{y})]=P[y\ne\hat{y}]$, where $E$ and $ P$ represent the expectation and the probability of its argument respectively. 
The distortion between sequences $\vec{y}, \hat{\vec{y}} \in {\cal Y}^M$ is defined by $d(\vec{y},\hat{\vec{y}})=\sum_{\mu=1}^M d(y^\mu,\hat{y}^\mu)$. 
Therefore, the distortion associated with the code is defined as $D=E[\frac 1M d(\vec{y},\hat{\vec{y}})]$, where the expectation is with respect to the probability distribution on ${\cal Y}$. 
A rate distortion pair $(R,D)$ is said to be {\it achievable} if there exists a sequence of rate distortion codes $({\cal F},{\cal G})$ with $E[\frac 1M d(\vec{y},\hat{\vec{y}})] \le D$ in the limit $M\to\infty$. 
We can now define a function to describe the boundary called the {\it rate distortion function}. 
The rate distortion function $R(D)$ is the infimum of rates $R$ such that $(R,D)$ is in the rate distortion region of the source for a given distortion $D$ and all rate distortion codes. 
The infimum of rates $R$ for a given distortion $D$ and given rate distortion codes $({\cal F},{\cal G})$ is called the {\it rate distortion property} of $({\cal F},{\cal G})$. 
We restrict ourselves to a Boolean source ${\cal Y}=\{0,1\}$. 
We assume that the source sequence is not biased to rule out the possibility of compression due to redundancy. 
The non-biased Boolean message in which each component is generated independently from an identical distribution $P(y^\mu=1)=P(y^\mu=0)=1/2$. 
For this simple source, the rate distortion function for an unbiased Boolean source with Hamming distortion is given by 
\begin{equation}
R(D)=1-h_2(D), 
\label{eq:RDF}
\end{equation}
where $h_2(x)=-x\log_2(x)-(1-x)\log_2(1-x)$ called the binary entropy function. 

\section{compression using multilayer perceptrons}

To simplify notations, let us replace all the Boolean representations $\{0,1\}$ with the Ising representation $\{1,-1\}$ throughout the rest of this paper. 
We set ${\cal Y}={\cal S}=\hat{\cal Y}=\{1,-1\}$ as the binary alphabets. 
We consider an unbiased source message in which a component is generated independently from an identical distribution: 
\begin{equation}
P(y^\mu)=\frac 12 \delta (y^\mu -1) + \frac 12 \delta (y^\mu +1), 
\end{equation}
for simplicity. 
First let us define a decoder. 
We can construct a nonlinear map ${\cal G} : {\cal S}^N \to \hat{\cal Y}^M$ from codeword $\vec{s}\in{\cal S}^N$ to reproduced message $\hat{\vec{y}}=(\hat{y}^\mu)\in\hat{\cal Y}^M$. 
For a given source message $\vec{y}=(y^\mu)\in{\cal Y}^M$, the role of the encoder is to find a codeword $\vec{s}\in{\cal S}^N$ that minimizes the distortion between its reproduced message ${\cal G}(\vec{s})$ and the source message $\vec{y}$. 
\par
\begin{figure}[t]
  \vspace{10mm}
  \begin{center}
  \includegraphics[width=.6\linewidth,keepaspectratio]{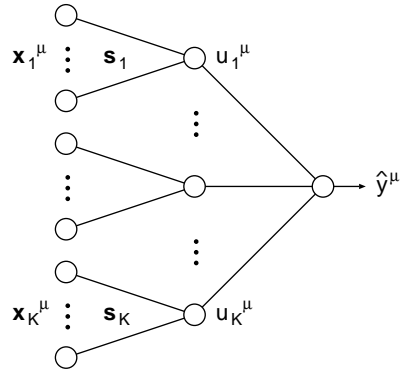}
  \end{center}
  \caption{The architecture of tree-like multilayer perceptrons with $N$ input units and $K$ hidden units. }
  \label{fig:tree}
  \vspace{10mm}
\end{figure}

We choose a nonlinear map ${\cal G}$ utilizing tree-like multilayer perceptrons, i.e., a tree-like committee machine (committee tree) and a tree-like parity machine (parity tree). 
Figure \ref{fig:tree} shows its architecture. 
The codeword $\vec{s}$ is divided into $N/K$-dimensional $K$ disjoint vectors $\vec{s}_1, \cdots ,\vec{s}_K\in{\cal S}^{N/K}$ as $\vec{s}={}^t(\vec{s}_1, \cdots, \vec{s}_K)$. 
The $l$th hidden unit receives the vector $\vec{s}_l$. 
The outputs of the committee tree and the parity tree are a majority decision and a parity of hidden unit outputs, respectively. 
The $\mu$th bit of the reproduced message $\hat{y}^\mu$ is defined by utilizing the committee tree as 
\begin{equation}
\hat{y}^\mu(\vec{s}) \equiv \sgn \biggl( \sum_{l=1}^K f \biggl( \sqrt{\frac KN}\vec{s}_l \cdot \vec{x}_l^\mu \biggr) \biggr), 
\label{eq:decoder_CM}
\end{equation}
where $\vec{x}_l^{\mu} \sim {\cal N}(\vec{0},\vec{1})$ are fixed $N/K$-dimensional vectors and the map $f:\mathbb{R}\to{\cal Y}$ is a transfer function. 
Function $\sgn(x)$ denotes the sign function taking 1 for $x \ge 0$ and -1 for $x<0$. 
Similarly, the $\mu$th bit $\hat{y}^\mu$ of the reproduced message is also defined by utilizing the parity tree as 
\begin{equation}
\hat{y}^\mu(\vec{s}) \equiv \prod_{l=1}^K f \biggl( \sqrt{\frac KN}\vec{s}_l \cdot \vec{x}_l^\mu \biggr). 
\label{eq:decoder_PM}
\end{equation}
The decoder ${\cal G}$ from the codeword $\vec{s}$ to the reproduced message $\hat{\vec{y}}=(\hat{y}^\mu)$ is described as 
\begin{equation}
{\cal G}(\vec{s}) \equiv \hat{\vec{y}}(\vec{s})={}^t(\hat{y}^1(\vec{s}),\cdots ,\hat{y}^M(\vec{s})), 
\end{equation}
In this framework, the encoder ${\cal F}$ from the original message $\vec{y}$ to the codeword $\vec{s}$ can be written as 
\begin{equation}
{\cal F}(\vec{y}) \equiv \mathop{\rm argmin}_{\hat{\vec{s}}} d(\vec{y},{\cal G}(\hat{\vec{s}})), 
\label{eq:encode}
\end{equation}
with respect to the case of both the committee tree and the parity tree. 
Employing the Ising representation, where the length of the codeword is infinite, the average Hamming distortion can be represented as 
\begin{equation}
E[d(\vec{y},\hat{\vec{y}})] = \sum_{\mu=1}^M [ 1- \Theta ( y^\mu \hat{y}^\mu) ], 
\end{equation}
where the function $\Theta(x)$ denotes the step function taking 1 for $x \ge 0$ and 0 otherwise. 
Since we assume the unbiased source message in this paper, we set $f(x)=\sgn(x)$. 
\par
This encoding scheme is essentially the same as a learning of the multilayer perceptrons because of a following reason. 
We first assign the random input vector $\vec{x}^\mu={}^t(\vec{x}_1^\mu,\cdots,\vec{x}_K^\mu) \in \mathbb{R}^N$ to each bit of the original message $y^\mu$. 
The encoder must find a weight vector $\vec{s}$ which satisfies input-output relations $\vec{x}^\mu \mapsto y^\mu$ as much as possible. 
Then we use this optimal weight vector $\vec{s}$ as a codeword. 
Therefore, in a lossless case of $D=0$, an evaluation of the rate distortion property of these codes is entirely identical to the calculation of the storage capacity \cite{Gardner1988,Krauth1989}. 

\section{analytical evaluation}

We analytically evaluate the typical performance, according to Hosaka et al \cite{Hosaka2002}, for the proposed compression scheme using the replica method. 
The minimum permissible average distortion $D$ is calculated, when the code rate $R$ is fixed. 
For a given original message $\vec{y}$ and the input vectors $\{\vec{x}_l^{\mu}\}$, the number of codewords $\vec{s}$, which provide a fixed Hamming distortion $MD=d(\vec{y},\hat{\vec{y}})$, can be expressed as 
\begin{equation}
{\cal N}(D,R) = \mathop{\rm Tr}_{\vec{s}} \delta \biggl( MD ; d(\vec{y},\hat{\vec{y}}(\vec{s})) \biggr), 
\end{equation}
where $\delta(m;n)$ denotes Kronecker's delta taking 1 if $m=n$ and 0 otherwise. 
Since the original message $\vec{y}$ and the input vectors $\{\vec{x}_l^\mu\}$ are randomly generated predetermined variables, the quenched average of the entropy per bit over these parameters, 
\begin{equation}
S(D,R)= \frac 1N <\ln {\cal N}(D,R)>_{\vec{y},\vec{x}}, 
\label{eq:def_S}
\end{equation}
is naturally introduced for investigating the typical properties, where $< \; >_{\vec{y},\vec{x}}$ denotes the average over $\vec{y}$ and $\{\vec{x}_l^{\mu}\}$. 
We calculate the entropy $S(D)$ by the replica method (see Appendix \ref{appendix.ReplicaMethod}). 
The rate-distortion region can be represented by $\{(D,R) | S(D,R) \ge 0 \}$. 
Therefore, a minimum code rate $R$ for a fixed distortion $D$ is given by a solution of $S(D,R)=0$. 
\par 
Note that a minimum code rate $R$ for $D=0$ coincides with a reciprocal of the critical storage capacity of a multilayer perceptron, i.e., the critical storage capacity $\alpha_c(\equiv M/N)$ can be obtained by $S(0,1/\alpha_c)=0$. 

\subsection{Replica symmetric theory of lossy compression using committee tree}

\subsubsection{For general $K$}

In the lossy compression using the committee tree, we obtain average entropy $S_{CT}(D,R)$ as 
\begin{eqnarray}
S_{CT}(D,R) 
&=& \mathop{\rm extr}_{\beta,q,\hat{q}} \biggl( R^{-1} \biggl< \int \biggl( \prod_{l=1}^K Dt_l \biggr) \nonumber \\
& & \times \ln \{ e^{-\beta} + (1-e^{-\beta}) \Sigma (\{t_l\};y) \} \biggr>_y \nonumber \\
& & + \int Du \ln 2 \cosh \sqrt{\hat{q}} u - \frac{\hat{q}(1-q)}2 \nonumber \\
& & + R^{-1}\beta D \biggr), 
\label{eq:S}
\end{eqnarray}
where 
\begin{equation}
\Sigma (\{t_l\};y) \equiv \mathop{\rm Tr}_{\{ \tau_l =\pm 1\}} \Theta \biggl( -y\sum_{l=1}^K \tau_l \biggr) \prod_{l=1}^K H(Q \tau_l t_l), 
\label{eq:Sigma}
\end{equation}
with $Q \equiv \sqrt{q/(1-q)}$ (see Appendix \ref{appendix.RM.CT.generalK}). 
For any $K$, we can obtain a minimum code rate $R$, which gives $S_{CT}(D,R)=0$ for a fixed distortion $D$.

\subsubsection{For large $K$}

We concentrate in the following on the simple case of large $K$, where the $K$-multiple integrals can be reduced to a single Gaussian integral. 
We assume that the number of hidden units $K$ is large but still $K \ll N$. 
Using the central limit theorem, the averaged entropy is given by 
\begin{eqnarray}
S_{CT}(D,R) 
&=& \mathop{\rm extr}_{\beta,q,\hat{q}} \biggl( R^{-1} \biggl< \int Dt \ln \{ e^{-\beta} \nonumber \\
& & + (1-e^{-\beta}) H \biggl( \sqrt{\frac{q_{eff}}{1-q_{eff}}}t \biggr) \} \biggr>_y \nonumber \\
& & + \int Du \ln 2 \cosh \sqrt{\hat{q}} u - \frac{\hat{q}(1-q)}2 \nonumber \\
& & + R^{-1}\beta D \biggr) , 
\label{eq:S_CT}
\end{eqnarray}
where $q_{eff} \equiv \int Dt [1-2H(Qt)]^2 = \frac 2{\pi} \sin^{-1} q$ 
and $Q_{eff} \equiv \sqrt{q_{eff}/(1-q_{eff})}$ (see Appendix \ref{appendix.RM.CT.largeK}). 
Figure \ref{fig:CT} shows that the limit of achievable code rate $R$ expected for $N \to \infty$ plotted versus the distortion $D$ for $K=1, 3$ and $K\to\infty$. 
For a fixed code rate $R$, the achievable distortion decreases as the number of hidden units $K$ increases. 
However, it does not saturate Shannon's limit even if in the limit $K\to\infty$. 
For large $K$, the EA order parameter $q$, which means the average overlap between different codewords, does not converge to zero. 
Since this means that codewords are correlated, the distribution of codewords is biased in ${\cal S}^N$. 
Note that a nonzero EA order parameter does not mean that the reproduced message has a nonzero average due to the random input vector which have a zero average.  
\begin{figure}[t]
  \vspace{2mm}
  \begin{center}
  \includegraphics[width=0.9\linewidth,keepaspectratio]{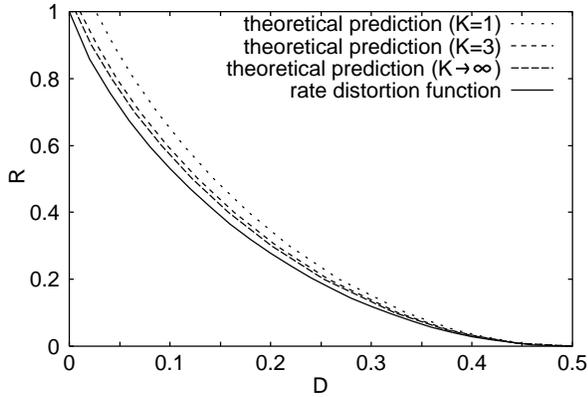}
  \end{center}
  \caption{
The rate distortion property of lossy compression using a committee tree. 
The limit of achievable code rate $R$ expected for $N \to \infty$ plotted versus the distortion $D$ for $K=1$ (dotted line), $K=3$ (short dashed line) and $K\to\infty$ (long dashed line). 
Solid line denotes rate-distortion function $R(D)$ for binary sequences by Shannon. 
}
  \label{fig:CT}
  \vspace{10mm}
\end{figure}

\subsection{Replica symmetric theory of lossy compression using parity tree}

In the lossy compression using the parity tree, on the other hand, we hence obtain averaged entropy $S_{PT}(D,R)$ as 
\begin{eqnarray}
S_{PT}(D,R) 
&=& \mathop{\rm extr}_{\beta,q,\hat{q}} \biggl( R^{-1} \biggl< \int \biggl( \prod_{l=1}^K Dt_l \biggr) \nonumber \\
& & \times \ln \{ e^{-\beta} + (1-e^{-\beta}) \Pi (\{t_l\};y) \} \biggr>_y \nonumber \\
& & + \int Du \ln 2 \cosh \sqrt{\hat{q}} u - \frac{\hat{q}(1-q)}2 \nonumber \\
& & + R^{-1}\beta D \biggr), 
\label{eq:S_PT}
\end{eqnarray}
where 
\begin{equation}
\Pi (\{t_l\};y) \equiv \frac 12 \biggl( 1+y\prod_{l=1}^K [1-2H(Q t_l)] \biggr) . 
\end{equation}
For cases utilizing a committee tree and a parity tree, only terms $\Sigma (\{t_l\};y)$ and $\Pi (\{t_l\};y)$ are different. 
Since both the order parameters $q$ and $\hat{q}$ at the saddle-point of (\ref{eq:S_PT}) are less than one, the average entropy can be expanded with respect to $\prod_{l=1}^K [1-2H(Q t_l)] (<1)$. 
Solutions of the saddle-point equation derived from the expanded form of average entropy are obtained as 
\begin{equation}
\left\{
\begin{array}{l}
q=0, \\
\hat{q}=0, \\
D=\displaystyle{\frac{e^{-\beta}}{1+e^{-\beta}}}, 
\end{array}
\right.
\label{eq:K>=2PTspe}
\end{equation}
in the case $K\ge 2$ (see Appendix \ref{appendix.RM.PT}). 
For $K=1$, $q>0$ holds. 
Note that for $K=1$, a parity tree is equivalent to a committee tree. 
For $K\ge 2$, the order parameter $q$ becomes zero, namely all codewords are uncorrelated and distributed all round in ${\cal S}^N$. 
Where $K \ge 2$, substituting (\ref{eq:K>=2PTspe}) into (\ref{eq:S_PT}), average entropy is obtained as 
\begin{eqnarray}
S_{PT}(D,R)
&=& -R^{-1}\ln 2+ \ln 2 -R^{-1}D\ln D \nonumber \\
& & -R^{-1}(1-D)\ln (1-D). 
\end{eqnarray}
A minimum code rate $R$ for a fixed distortion $D$ and $K \ge 2$ is given by $S_{PT}(D,R)=0$. 
Solving this equation with respect to $R$, we obtain 
\begin{equation}
R = 1-h_2(D) \equiv R_{RS}(D), 
\end{equation}
which is identical to the rate-distortion function for uniformly unbiased binary sources (\ref{eq:RDF}). 
\par
\begin{figure}[t]
  \vspace{2mm}
  \begin{center}
  \includegraphics[width=0.9\linewidth,keepaspectratio]{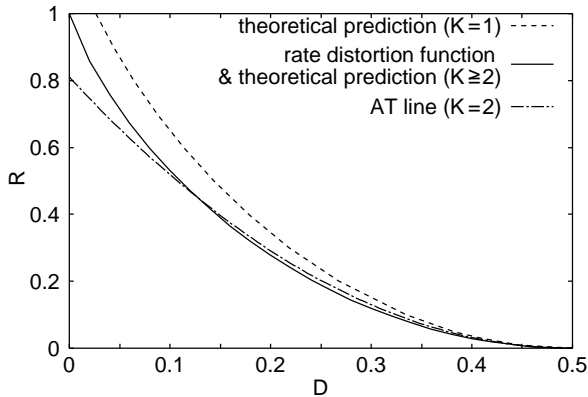}
  \end{center}
  \caption{
The rate distortion property of lossy compression using a parity tree. 
The limits of achievable code rate $R$ expected for $N \to \infty$ plotted versus the distortion $D$ for $K=1$ (dashed line) and $K \ge 2$ (solid line). 
Solid line also denotes rate-distortion function, which is identical to limit of achievable distortion for $K \ge 2$. 
Dash-dotted line denotes AT line for $K=2$. 
For $K\ge 3$, RS solution does not exhibit AT instability throughout achievable region. 
}
  \label{fig:PT}
  \vspace{10mm}
\end{figure}

However, since calculation is based on the RS ansatz, we verify the AT stability to confirm the validity of this solution. 
As the RS solution to lossy compression using a parity tree with $K=2$ hidden units can be simply expressed as (\ref{eq:K>=2PTspe}), the stability condition is analytically obtained as 
\begin{equation}
R>\frac 8{\pi^2}(1-2D)^2 \equiv  R_{AT}(D), 
\label{eq:ATline}
\end{equation}
where boundary $R=R_{AT}(D)$ is called the AT line (see Appendix \ref{appendix.ATline}). 
For $K\ge 3$, the RS solution does not exhibit the AT instability throughout the achievable region of the rate-distortion pair $(R,D)$. 
Figure \ref{fig:PT} shows the limit of achievable distortion $D$ expected for $N \to \infty$ plotted versus code rate $R$ for $K=1$ and $K\ge 2$. 
In the case $K\ge 2$, the limit of achievable distortion is identical to the rate-distortion function. 
The dash-dotted line in Fig. \ref{fig:PT} denotes the AT line for $K=2$. 
The region above the AT line denotes that the RS solution is stable. 
For $K=2$, we found that for distortion $0.126 \alt D \le 0.5$, $R_{RS}(D)$ can become smaller than $R_{AT}(D)$. 
Nevertheless this instability may not be serious in practice, because the region where the RS solution becomes unstable is narrow. 
\par 
The annealed approximation of the entropy (\ref{eq:def_S}) gives a lower bound to the rate distortion property. 
It coincides with the rate distortion function. 
According to Opper's discussion \cite{Opper1995}, the entropy (\ref{eq:def_S}) can be represented by the information entropy formally. 
The annealed information entropy can give a upper bound to the rate distortion property. 
However, its evaluation is difficult (see Appendix \ref{appendix.lower_bound}). 

\section{distribution of codewords}

\begin{figure}[t]
  \vspace{2mm}
  \begin{center}
  \includegraphics[width=.8\linewidth,keepaspectratio]{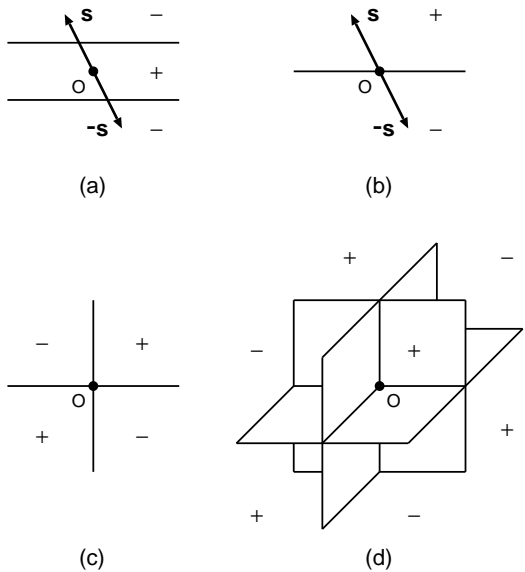}
  \end{center}
  \caption{
Relationship between codeword and bit of reproduced message in lossy compression using parity tree with $K$ hidden units. 
Symbol $+$ denotes bit of the reproduced message is $1$ and $-$ denotes that it is $-1$. 
Set ${\cal S}^N$ is divided by $K$ hyperplanes, whose normal vectors are orthogonal each other. 
For $K\ge 2$, vectors with same distortion as codeword $\vec{s}$ are distributed throughout ${\cal S}^N$. 
(a) a nonmonotonic perceptron, $q=0$, 
(b) a $K=1$ parity tree, $q>0$, 
(c) a $K=2$ parity tree, $q=0$, and 
(d) a $K=3$ parity tree, $q=0$.
}
  \label{fig:distribution}
  \vspace{10mm}
\end{figure}

It has already been shown that both compression using a sparse matrix and compression using a nonmonotonic perceptron also achieve optimal performance known as Shannon's limit \cite{Murayama2003,Hosaka2002}. 
All these schemes and compression using a parity tree with $K\ge 2$ hidden units becomes the common EA order parameter $q=0$. 
In compression using a nonmonotonic perceptron, 
the $\mu$th bit of the reproduced message is defined as $\hat{y}^\mu(\vec{s}) \equiv \hat{f} ( N^{-1/2} \vec{s} \cdot \vec{x}^\mu )$, where $\hat{f}$ is the transfer function with mirror symmetry, i.e., $\hat{f}(-x)=\hat{f}(x)$ \cite{Hosaka2002}. 
Due to the mirror symmetry of $\hat{f}$, both $\vec{s}$ and $-\vec{s}$ provide identical output for any $\vec{x}^\mu$. 
Hence, the EA order parameter is likely to become zero. 
The transfer function $\hat{f}$ with parameter $\kappa$ is defined as taking 1 for $|x| \le \kappa$ and $-1$ otherwise. 
Figure \ref{fig:distribution} shows the relationship between a codeword and a bit of the reproduced message. 
Figure \ref{fig:distribution} (a) is the case of compression using a nonmonotonic perceptron. 
\par
In compression using a parity tree, on the other hand, the $\mu$th bit of the reproduced message is 
\begin{equation}
\hat{y}^\mu(-\vec{s})
= \prod_{l=1}^K {\rm sgn} \biggl( \sqrt{\frac KN} \vec{x}_l^{\mu} \cdot (-\vec{s}_l)\biggr)
= (-1)^K \hat{y}^\mu(\vec{s}). 
\label{eq:mirror}
\end{equation}
For $K=1$, i.e., a parity tree is identical to a monotonic perceptron, $\hat{y}^\mu(-\vec{s}) = - \hat{y}^\mu(\vec{s})$ holds. 
Here, the EA order parameter becomes $q>0$. 
Therefore, the distribution of codewords is biased in ${\cal S}^N$. 
Compression using a parity tree with $K=1$ hidden unit cannot achieve Shannon's limit. 
Figure \ref{fig:distribution} (b) shows the case of compression using a monotonic perceptron, i.e., a committee tree and a $K=1$ parity tree. 
However, for an even number of hidden units $K$, a parity tree also has the same effect as mirror symmetry. 
\par
We will next discuss the case of $K\ge 2$. 
Let ${\cal V}(\vec{s}) \subset {\cal S}^N$ be a set of vectors that reversed the signs of an arbitrary even number of blocks of a codeword $\vec{s}={}^t(\vec{s}_1, \cdots , \vec{s}_K)$, e.g., ${}^t(-\vec{s}_1, -\vec{s}_2 ,\vec{s}_3, \cdots , \vec{s}_K) \in {\cal V}(\vec{s})$. 
The cardinality of the set ${\cal V}(\vec{s})$ is 
\begin{equation}
||{\cal V}(\vec{s})|| = \sum_{n=0}^{\lfloor K/2 \rfloor} {}_K C_{2n}=2^{K-1},
\end{equation}
where $\lfloor x \rfloor$ is the largest integer $\le x$. 
According to (\ref{eq:decoder_PM}), all $\hat{\vec{s}} \in {\cal V}(\vec{s})$ provide identical output for any $\vec{x}_l^\mu$. 
The summation of all $\hat{\vec{s}} \in {\cal V}(\vec{s})$ becomes 
\begin{equation}
\sum_{\hat{\vec{s}} \in {\cal V}(\vec{s})} \hat{\vec{s}}
= {}^t (\cdots , 2^{K-2}\vec{s}_l + 2^{K-2}(-\vec{s}_l),\cdots ) =\vec{0}. 
\end{equation}
This means that $2^{K-1}$ vectors with the same distortion as codeword $\vec{s}$ are distributed throughout ${\cal S}^N$. 
For instance, Fig. \ref{fig:distribution} (c) shows the distribution of codewords obtained by compression using a $K=2$ parity tree. 
The set ${\cal S}^N$ is divided by two $N-1$-dimensional hyperplanes whose normal vectors are orthogonal to each other. 
For the $\mu$th bit of the reproduced message, the normal vectors of hyperplanes are ${}^t(\vec{x}_1^\mu,\vec{0})$ and ${}^t(\vec{0},\vec{x}_2^\mu) \in \mathbb{R}^N$. 
Figure \ref{fig:distribution} (d) shows the case of compression using a $K=3$ parity tree. 
Here, although the same effect as mirror symmetry cannot be seen, nevertheless, EA order parameter $q$ becomes zero for the reason mentioned above. 
This situation is the same for $K \ge 4$. 
\par 
With respect to LDGM code \cite{Murayama2003}, Murayama succeeded in developing a practical encoder using the Thouless-Anderson-Palmer (TAP) approach which introduced inertia term heuristically \cite{Murayama2004}. 
The TAP approach is called belief propagation (BP) in the field of information theory. 
Hosaka et al applied this inertia term introduced BP to compression using a nonmonotonic perceptron \cite{Hosaka2005b}. 
In compression using a parity tree with $K$ hidden units, the number of codewords which give a minimun distortion is $2^{K-1}$. 
Therefore, it may become easy to find codewords as the number of hidden units $K$ becomes large. 
But, in a practical encoding problem, it may not be easy to use a large $K$ since $K \ll N$ is needed. 

\section{conclusion}

We investigated a lossy compression scheme for unbiased Boolean messages employing a committee tree and a parity tree, whose transfer functions were monotonic. 
The lower bound for achievable distortion in using a committee tree became small when the number of hidden units $K$ was large. 
It did not reach Shannon's limit even in the case where $K\to\infty$. 
However, lossy compression using a parity tree with $K \ge 2$ hidden units could achieve Shannon's limit where the code length became infinity. 
We assumed the RS ansatz in our analysis using the replica method. 
In using a parity tree with $K \ge 2$, the RS solution was unstable in the narrow region. 
For $K\ge 3$, the RS solution did not exhibit the AT instability throughout the achievable region. 
\par
There is generally more than one code with the same distortion as a codeword. 
The EA order parameter, which means an average overlap between different codewords, need to be zero to reach Shannon's limit like several known schemes which saturate this limit. 
Therefore, it may be a necessary condition that the EA order parameter becomes zero to reach Shannon's limit. 
\par
Since the encoding with our method needs exponential-time, we need to employ various efficient polynominal-time approximation encoding algorithms. 
It is under way to investigate the influence of the number of hidden units on the accuracy of approximation encoding algorithms. 
In future work, we intend to evaluate the upper bound to the rate distortion property without replica. 

\vspace*{5mm}
\begin{center}
{\small \bf ACKNOWLEDGEMENTS}
\end{center}

This work was partially supported by a Grant-in-Aid 
for Scientific Research on Priority Areas No. 14084212, 
and for Scientific Research (C) No. 16500093, and 
for Encouragement of Young Scientists (B) No. 15700141 
from the Ministry of Education, Culture, Sports, Science and Technology of Japan.

\vspace*{5mm}
\appendix

\section{Analytical Evaluation using the replica method}
\label{appendix.ReplicaMethod}

The entropy $S(D,R)$ can be evaluated by the replica method: 
\begin{equation}
S(D,R) = \lim_{n \to 0} \frac 1{nN} \ln <{\cal N}^n (D,R)>_{\vec{y},\vec{x}}. 
\end{equation}
A moment ${\cal N}^n(D,R)$, which is the number of codewords with respect to an $n$-replicated system, can be represented as 
\begin{equation}
{\cal N}^n(D,R) = \mathop{\rm Tr}_{\vec{s}^1,\cdots ,\vec{s}^n} \prod_{a=1}^n \delta \biggl( MD ; d(\vec{y},\hat{\vec{y}}(\vec{s}^a)) \biggr), 
\end{equation}
where $\vec{s}^a ={}^t(\vec{s}_1^a,\cdots ,\vec{s}_K^a)$ and the superscript $a$ denotes a replica index. 
Inserting an identity 
\begin{eqnarray}
1
&=& \prod_{a<b} \prod_{l=1}^K \int_{-\infty}^\infty dq_l^{ab} \delta \biggl( \vec{s}_l^a \cdot \vec{s}_l^b -\frac NK q_l^{ab} \biggr) \nonumber \\
&=& \biggl( \frac 1{2\pi i} \biggr)^{n(n-1)K/2} \int \biggl( \prod_{a<b} \prod_l dq_l^{ab} d\hat{q}_l^{ab} \biggr) \nonumber \\
& & \times \exp \biggl[ \sum_{a<b} \sum_l \hat{q}_l^{ab} \biggl(\vec{s}_l^a \cdot \vec{s}_l^b -\frac NK q_l^{ab} \biggr) \biggr] ,
\end{eqnarray}
into this expression to separate the relevant order parameter. 
Utilizing the Fourier expression of Kronecker's delta, 
\begin{align}
\delta \biggl( MD;d(\vec{y},\hat{\vec{y}}(\vec{s}^a)) \biggr) =& \int_{i(c-\pi)}^{i(c+\pi)} \frac{d\beta^a}{2\pi i} e^{\beta^a(D-d(\vec{y},\hat{\vec{y}}(\vec{s}^a)))}, \nonumber \\
&\qquad \qquad \forall c \in \mathbb{R},
\end{align}
we can calculate the average moment $<{\cal N}^n (D,R)>_{\vec{y},\vec{x}}$ for natural numbers $n$ as 
\begin{widetext}
\begin{eqnarray}
<{\cal N}^n (D,R)>_{\vec{y},\vec{x}} 
&\simeq& \int \biggl( \prod_a d\beta^a \biggr) \int \biggl( \prod_{a<b} \prod_l dq_l^{ab} d\hat{q}_l^{ab} \biggr) \exp N \biggl[ \nonumber \\
& & R^{-1} \ln \biggl< \int \biggl( \prod_l d\vec{u}_l d\vec{v}_l \biggr) \prod_l e^{-\frac 12 {}^t\vec{v}_l Q_l \vec{v}_l + i\vec{v}_l\cdot\vec{u}_l} \prod_a \biggl\{ e^{-\beta^a} + (1-e^{-\beta^a}) \Theta ( y,\{u_l^a\}) \biggr\} \biggr>_y \nonumber \\
& & + \frac 1K \ln \mathop{\rm Tr}_{\{s_l^a\}} \exp \biggl( \sum_{a<b} \sum_l \hat{q}_l^{ab} s_l^a s_l^b \biggr) - \frac 1K \sum_{a<b} \sum_l q_l^{ab} \hat{q}_l^{ab} + R^{-1} D \sum_a \beta^a \biggr], 
\label{eq:Nn}
\end{eqnarray}
\end{widetext}
where $Q_l$ is an $n \times n$ matrix having matrix elements $\{ q_l^{ab} \}$ and $< h(y) >_y = \sum_{y\in\{-1,1\}} [\frac 12\delta(y-1)+\frac 12\delta(y+1)]h(y)$. 
Function $\Theta ( y,\{u_l^a\})$ included in the right hand side of (\ref{eq:Nn}) depends on the decoder (details are discussed in the following sections). 
We analyze a system in the thermodynamic limit $N,M \to \infty$, while code rate $R$ is kept finite. 
This integral (\ref{eq:Nn}) will be dominated by the saddle-point of the extensive exponent and can be evaluated via a saddle-point problem with respect to $\beta^a, q_l^{ab}$ and $\hat{q}_l^{ab}$. 
Here, we assume the replica symmetric (RS) ansatz: 
\begin{equation}
\left\{
\begin{array}{l}
\beta_a = \beta, \\
q_l^{ab} = (1-q) \delta_{ab} + q, \\
\hat{q}_l^{ab} = (1-\hat{q}) \delta_{ab} + \hat{q}, 
\end{array}
\right.
\end{equation}
where $\delta_{k,k'}$ is Kronecker's delta taking 1 if $k=k'$ and 0 otherwise. 
This ansatz means that all the hidden units are equivalent after averaging over the disorder.

\subsection{Lossy compression using committee tree \\ for general $K$}
\label{appendix.RM.CT.generalK}

In the lossy compression using the committee tree, the $\Theta ( y,\{u_l^a\})$ included in (\ref{eq:Nn}) is obtained as 
\begin{equation}
\Theta ( y,\{u_l^a\}) = \Theta \biggl( y\sum_{l=1}^K \sgn (u_l^a) \biggr). 
\end{equation}
Therefore, we obtain average entropy $S_{CT}(D,R)$ as 
\begin{eqnarray}
S_{CT}(D,R) 
&=& \mathop{\rm extr}_{\beta,q,\hat{q}} \biggl( R^{-1} \biggl< \int \biggl( \prod_{l=1}^K Dt_l \biggr) \nonumber \\
& & \times \ln \{ e^{-\beta} + (1-e^{-\beta}) \Sigma (\{t_l\};y) \} \biggr>_y \nonumber \\
& & + \int Du \ln 2 \cosh \sqrt{\hat{q}} u - \frac{\hat{q}(1-q)}2 \nonumber \\
& & + R^{-1}\beta D \biggr), 
\label{eq:app.S}
\end{eqnarray}
where 
\begin{equation}
\Sigma (\{t_l\};y) \equiv \mathop{\rm Tr}_{\{ \tau_l =\pm 1\}} \Theta \biggl( -y\sum_{l=1}^K \tau_l \biggr) \prod_{l=1}^K H(Q \tau_l t_l), 
\label{eq:app.Sigma}
\end{equation}
with $Q \equiv \sqrt{q/(1-q)}$. 
Utilizing the Fourier expression of the step function $\Theta(x)=\int_0^\infty d\lambda \int_{-i\infty}^{i\infty} \frac{d\hat{\lambda}}{2\pi i}e^{\hat{\lambda}(\lambda -x)}$, 
the saddle-point equations $\frac{\partial S}{\partial \beta}=\frac{\partial S}{\partial q}=\frac{\partial S}{\partial \hat{q}}=0$ become 
\begin{eqnarray}
q &=& \int Du \tanh^2 \sqrt{\hat{q}} u, \\
\hat{q} &=& 2R^{-1} \biggl< \int \biggl( \prod_{l=1}^K Dt_l \biggr) \frac{-(1-e^{-\beta})\Sigma '(\{ t_l \};y)}{e^{-\beta}+(1-e^{-\beta})\Sigma (\{ t_l \};y)} \biggr>_y, \nonumber \\
& & \\
D &=& \biggl< \int \biggl( \prod_{l=1}^K Dt_l \biggr) \frac{e^{-\beta}-e^{-\beta}\Sigma (\{ t_l \};y)}{e^{-\beta}+(1-e^{-\beta})\Sigma (\{ t_l \};y)} \biggr>_y, 
\end{eqnarray}
where $\Sigma '(\{t_l\};y) \equiv \partial \Sigma (\{t_l\};y)/\partial q$. 
Substituting the solutions to the saddle-point equations into (\ref{eq:app.S}), average entropy $S_{CT}(D,R)$ is obtained. 
Thus, for any $K$, we can obtain a minimum code rate $R$, which gives $S_{CT}(D,R)=0$ for a fixed distortion $D$.

\subsection{Lossy compression using committee tree \\ for large $K$}
\label{appendix.RM.CT.largeK}

We concentrate in the following on the simple case of large $K$, where the $K$-multiple integrals can be reduced to a single Gaussian integral. 
We assume that the number of hidden units $K$ is large but still $K \ll N$. 
Here, the term $\Sigma (\{t_l\};y)$ included in (\ref{eq:app.S}) does not depend on all the individual integration variables $t_l$ but only on the combination $\sum_{l=1}^K [2H(Qt_l)-1]$. 
With the central limit theorem, the term is given by 
\begin{eqnarray}
\Sigma (\{t_l\};y) &=& \int_0^\infty d\lambda \int_{-\infty}^\infty \frac{d\hat{\lambda}}{2\pi} \exp \biggl\{ i \hat{\lambda} \lambda \nonumber \\
& & + i \hat{\lambda} y \frac 1{\sqrt{K}}\sum_l [2H(Qt_l)-1] \nonumber \\
& & -\hat{\lambda}^2 \biggl( 1- \frac 1K \sum_l [2H(Qt_l)-1]^2 \biggr) \biggr\}. \nonumber \\
\end{eqnarray}
Therefore, we obtain averaged entropy as 
\begin{eqnarray}
S_{CT}(D,R) 
&=& \mathop{\rm extr}_{\beta,q,\hat{q}} \biggl( R^{-1} \biggl< \int Dt \ln \{ e^{-\beta} \nonumber \\
& & + (1-e^{-\beta}) H \biggl( \sqrt{\frac{q_{eff}}{1-q_{eff}}}t \biggr) \} \biggr>_y \nonumber \\
& & + \int Du \ln 2 \cosh \sqrt{\hat{q}} u - \frac{\hat{q}(1-q)}2 \nonumber \\
& & + R^{-1}\beta D \biggr) , 
\label{eq:app.S_CT}
\end{eqnarray}
where $q_{eff} \equiv \int Dt [1-2H(Qt)]^2 = \frac 2{\pi} \sin^{-1} q$ and the saddle-point equations are 
\begin{eqnarray}
q &=& \int Du \tanh^2 \sqrt{\hat{q}} u, \\
\hat{q} &=& 2R^{-1} \biggl< \int Dt \frac{-(1-e^{-\beta}) H'(Q_{eff}t)}{e^{-\beta}+(1-e^{-\beta})H(Q_{eff}t)} \biggr>_y, \\
D &=& \biggl< \int Dt \frac{e^{-\beta}-e^{-\beta}H(Q_{eff}t)}{e^{-\beta}+(1-e^{-\beta})H(Q_{eff}t)} \biggr>_y, 
\end{eqnarray}
with $Q_{eff} \equiv \sqrt{q_{eff}/(1-q_{eff})}$ and $H'(Q_{eff}t) \equiv \partial H(Q_{eff}t)/\partial q$.

\subsection{Lossy compression using parity tree \\ for general $K$}
\label{appendix.RM.PT}

In the lossy compression using the parity tree, on the other hand, the $\Theta ( y,\{u_l^a\})$ included in (\ref{eq:Nn}) is obtained as 
\begin{equation}
\Theta ( y,\{u_l^a\}) = \Theta \biggl( y\prod_l \sgn (u_l^a) \biggr), 
\end{equation}
Hence, we obtain averaged entropy $S_{PT}(D,R)$ as 
\begin{eqnarray}
S_{PT}(D,R) 
&=& \mathop{\rm extr}_{\beta,q,\hat{q}} \biggl( R^{-1} \biggl< \int \biggl( \prod_{l=1}^K Dt_l \biggr) \nonumber \\
& & \times \ln \{ e^{-\beta} + (1-e^{-\beta}) \Pi (\{t_l\};y) \} \biggr>_y \nonumber \\
& & + \int Du \ln 2 \cosh \sqrt{\hat{q}} u - \frac{\hat{q}(1-q)}2 \nonumber \\
& & + R^{-1}\beta D \biggr), 
\label{eq:app.S_PT}
\end{eqnarray}
where 
\begin{equation}
\Pi (\{t_l\};y) \equiv \frac 12 \biggl( 1+y\prod_{l=1}^K [1-2H(Q t_l)] \biggr) . 
\end{equation}
For cases utilizing a committee tree and a parity tree, only terms $\Sigma (\{t_l\};y)$ and $\Pi (\{t_l\};y)$ are different. 
Since both the order parameters $q$ and $\hat{q}$ at the saddle-point of (\ref{eq:app.S_PT}) are less than one, the average entropy $S_{PT}(D,R)$ can be expanded with respect to $\prod_{l=1}^K [1-2H(Q t_l)] (<1)$ as 
\begin{eqnarray}
S_{PT}(D,R)
&=& \mathop{\rm extr}_{\beta,q,\hat{q}} \biggl( R^{-1} \biggl\{ \ln \frac {1+e^{-\beta}}2 \nonumber \\
& & - \sum_{m=1}^\infty \frac 1{2m} \biggl( \frac {1-e^{-\beta}}{1+e^{-\beta}} \biggr)^{2m} \nonumber \\
& & \times \biggl[ \int Dt [1-2H(Qt)]^{2m} \biggr]^K \biggr\} \nonumber \\
& & + \int Du \ln 2 \cosh \sqrt{\hat{q}} u - \frac{\hat{q}(1-q)}2 \nonumber \\
& & + R^{-1}\beta D \biggr). 
\end{eqnarray}
We obtain saddle-point equations using this expanded form of the averaged entropy: 
\begin{eqnarray}
q
&=& \int Du \tanh^2 \sqrt{\hat{q}} u, \label{eq:app.q.PM} \\
\hat{q}
&=& 2R^{-1} K \sum_{m=1}^\infty \biggl( \frac{1-e^{-\beta}}{1+e^{-\beta}} \biggr)^{2m} \nonumber \\
& & \times \biggl[ \int Dt (1-2H(Qt))^{2m} \biggr]^{K-1} \nonumber \\
& & \times \int Dt (1-2H(Qt))^{2m-1} \frac{te^{-(Qt)^2/2}}{\sqrt{2\pi q}(1-q)^{3/2}}, \nonumber \\
& & \label{eq:app.q^.PM} \\
D
&=& \frac{e^{-\beta}}{1+e^{-\beta}} + \sum_{m=1}^\infty \frac{2e^{-\beta}}{(1+e^{-\beta})^2} \biggl( \frac{1-e^{-\beta}}{1+e^{-\beta}} \biggr)^{2m-1} \nonumber \\
& & \times \biggl[ \int Dt (1-2H(Qt))^{2m} \biggr]^K. \label{eq:app.D.PM} 
\end{eqnarray}
For $K\ge 2$, because of the existence of term $[ \int Dt (1-2H(Qt))^{2m} ]^{K-1}$ in (\ref{eq:app.q^.PM}), solutions to the saddle-point equations can become $q=\hat{q}=0$. 
We can find no other solutions except for $q=\hat{q}=0$ by solving (\ref{eq:app.q.PM})-(\ref{eq:app.D.PM}) numerically for $K\ge 2$. 
Substituting this into (\ref{eq:app.D.PM}), we obtain $D=e^{-\beta}/(1+e^{-\beta})$.

\section{Almeida-Thouless instability of replica symmetric solution} 
\label{appendix.ATline}

\subsection{General case}
\label{appendix.ATline.general}

The Hessian computed at the replica symmetric saddle-point characterizes fluctuations in the order parameters $\beta^a$, $q_l^{ab}$ and $\hat{q}_l^{ab}$ around the RS saddle-point. 
Instability of the RS solution is signaled by a change of sign of at least one of the eigenvalues of the Hessian. 
Let ${\cal M}(\{\beta^a\},\{ q_l^{ab} \},\{\hat{q}_l^{ab}\})$ be the exponent of the integrand of the integral (\ref{eq:Nn}). 
Equation (\ref{eq:Nn}) can be represented as 
\begin{eqnarray}
& & <{\cal N}^n(D,R)>_{\vec{y},\vec{x}} \nonumber \\
&=& \int \biggl( \prod_a d\beta^a \biggr) \int \biggl( \prod_{a<b} \prod_l dq_l^{ab} d\hat{q}_l^{ab} \biggr) \nonumber \\
& & \times \exp ( N {\cal M}(\{\beta^a\},\{ q_l^{ab} \},\{\hat{q}_l^{ab}\}) ). 
\end{eqnarray}
We expand ${\cal M}$ around $\beta$, $q$ and $\hat{q}$ in $\delta\beta^a$, $\delta q_l^{ab}$ and $\delta \hat{q}_l^{ab}$ and then find up to second order 
\begin{eqnarray}
& & {\cal M}(\{\beta+\delta\beta^a\},\{ q+q_l^{ab} \},\{\hat{q}+\delta\hat{q}_l^{ab}\}) \nonumber \\
&=& {\cal M}(\{\beta\},\{q\},\{\hat{q}\}) + \frac 12 {}^t\vec{\nu} G \vec{\nu} + {\cal O}(||\vec{\nu}||^3), 
\end{eqnarray}
where 
\begin{equation}
\vec{\nu}={}^t(\{ \delta \beta^a \} , \{ \delta q_1^{ab} \} , \{ \delta \hat{q}_1^{ab} \} , \cdots , \{ \delta q_K^{ab} \} , \{ \delta \hat{q}_K^{ab} \} ), 
\end{equation}
is the perturbation to the RS saddle-point. 
The Hessian $G$ is the following $[n+Kn(n-1)] \times [n+Kn(n-1)]$ matrix: 
\begin{equation}
G=
\left(
\begin{array}{ccccc}
S      & T      & T      & \cdots & T      \\
{}^tT  & U      & V      & \cdots & V      \\
{}^tT  & V      & U      & \cdots & V      \\
\vdots & \vdots & \vdots & \ddots & \vdots \\
{}^tT  & V      & V      & \cdots & U
\end{array}
\right), 
\end{equation}
where $n \times n$ matrix $S$, $n \times n(n-1)$ matrix $T$ and $n(n-1) \times n(n-1)$ matrices $U, V$ are 
\begin{eqnarray}
S &=& (\{S^{a,b}\}), \nonumber \\
T &=& (\{T^{a,bc}\},\{\hat{T}^{a,bc}\}), \nonumber \\
U &=&
\left(
\begin{array}{cc}
\{U^{ab,cd}\} & \{\tilde{U}^{ab,cd}\} \\
\{\tilde{U}^{ab,cd}\} & \{\hat{U}^{ab,cd}\}
\end{array}
\right), \nonumber \\
V &=&
\left(
\begin{array}{cc}
\{V^{ab,cd}\} & \{\tilde{V}^{ab,cd}\} \\
\{\tilde{V}^{ab,cd}\} & \{\hat{V}^{ab,cd}\}
\end{array}
\right), 
\end{eqnarray}
with 
\begin{eqnarray}
S^{a,b} &=& \partial^2 {\cal M}/\partial \beta^a \partial \beta^b, \nonumber \\
T^{a,bc} &=& \partial^2 {\cal M}/\partial \beta^a \partial q_l^{bc}, \nonumber \\
\hat{T}^{a,bc} &=& \partial^2 {\cal M}/\partial \beta^a \partial \hat{q}_l^{bc}, \nonumber \\
U^{ab,cd} &=& \partial^2 {\cal M}/\partial q_l^{ab} \partial q_l^{cd}, \nonumber \\
\hat{U}^{ab,cd} &=& \partial^2 {\cal M}/\partial \hat{q}_l^{ab} \partial \hat{q}_l^{cd}, \nonumber \\
\tilde{U}^{ab,cd} &=& \partial^2 {\cal M}/\partial q_l^{ab} \partial \hat{q}_l^{cd}, \nonumber \\
V^{ab,cd} &=& \partial^2 {\cal M}/\partial q_l^{ab} \partial q_{l'}^{cd} \quad (l\ne l'), \nonumber \\
\hat{V}^{ab,cd} &=& \partial^2 {\cal M}/\partial \hat{q}_l^{ab} \partial \hat{q}_{l'}^{cd} \quad (l\ne l'), \nonumber \\
\tilde{V}^{ab,cd} &=& \partial^2 {\cal M}/\partial q_l^{ab} \partial \hat{q}_{l'}^{cd} \quad (l\ne l'). 
\end{eqnarray}
For $(\beta,q,\hat{q})$ to be a local maximum of ${\cal M}$, it is necessary for the Hessian $G$ to be negative definite, i.e., all of its eigenvalues must be negative. 
Matrices $U$ and $V$ contain the quadratic fluctuations of the order parameters in the same and different hidden units, respectively. 
Because of the block form of $G$, the eigenproblem splits into an uncoupled diagonalization of the two matrices: $U-V$ and 
\begin{equation}
\hat{G}=
\left(
\begin{array}{ccc}
S      & \; & T \\
K{}^tT & \; & U+(K-1)V
\end{array}
\right).
\end{equation}
The eigenvectors of $U-V$ correspond to fluctuations in directions that break the permutation symmetry (PS). 
The eigenvectors of $\hat{G}$ represent fluctuations that do not break this symmetry. 
The most unstable mode corresponds to an eigenvector of $\hat{G}$ that breaks the replica symmetry (RS). 
We can write the eigenvalue equation as 
\begin{equation}
\hat{G} \vec{\mu} = \lambda \vec{\mu}, 
\end{equation}
with 
\begin{equation}
\vec{\mu}={}^t(\{ \epsilon^a \} , \{ \eta^{ab} \} , \{ \hat{\eta}^{ab} \} ). 
\end{equation}
There are three types of eigenvectors, i.e., $\vec{\mu}_1$, $\vec{\mu}_2$ and $\vec{\mu}_3$ \cite{Almeida1978}. 
The first $\vec{\mu}_1$ has the form: 
\begin{equation}
\epsilon^a=\epsilon, \quad \eta^{ab}=\eta, \quad \hat{\eta}^{ab}=\hat{\eta}. 
\end{equation}
Using the orthogonality of $\vec{\mu}_1$ and $\vec{\mu}_2$, the second type of eigenvector $\vec{\mu}_2$ has the form: 
\begin{eqnarray}
\epsilon^a &=&
\left\{
\begin{array}{ll}
(1-n) \epsilon, & (a=\theta), \\
\epsilon, & ({\rm otherwise}),
\end{array}
\right. \nonumber \\
\eta^{ab} &=& 
\left\{
\begin{array}{ll}
\frac 12(2-n) \eta, & (a=\theta \;{\rm or}\; b=\theta), \\
\eta, & ({\rm otherwise}),
\end{array}
\right. \nonumber \\
\hat{\eta}^{ab} &=& 
\left\{
\begin{array}{ll}
\frac 12(2-n) \hat{\eta}, & (a=\theta \;{\rm or}\; b=\theta), \\
\hat{\eta}, & ({\rm otherwise}),
\end{array}
\right. ,
\end{eqnarray}
for a specific replica $\theta$. 
In the limit $n\to 0$ this eigenvector $\vec{\mu}_2$ converges to $\vec{\mu}_1$, therefore the eigenvalue of the eigenvector $\vec{\mu}_2$ becomes degenerate with $\vec{\mu}_1$'s. 
\par
Similarly, using the orthogonality of $\vec{\mu}_2$ and $\vec{\mu}_3$, the third type of eigenvector $\vec{\mu}_3$ has the form: 
\begin{eqnarray}
\epsilon^a &=& 0, \nonumber \\
\eta^{ab} &=& 
\left\{
\begin{array}{l}
\frac 12(2-n)(3-n) \eta, \; (a=\theta,b=\nu), \\
\frac 12(3-n) \eta, \\
\quad (a=\theta \;{\rm or}\; a=\nu \;{\rm or}\; b=\theta \;{\rm or}\; b=\nu), \\
\eta, \; ({\rm otherwise}),
\end{array}
\right. \nonumber \\
\hat{\eta}^{ab} &=& 
\left\{
\begin{array}{l}
\frac 12(2-n)(3-n) \hat{\eta}, \; (a=\theta,b=\nu), \\
\frac 12(3-n) \hat{\eta}, \\
\quad (a=\theta \;{\rm or}\; a=\nu \;{\rm or}\; b=\theta \;{\rm or}\; b=\nu), \\
\hat{\eta}, \; ({\rm otherwise}),
\end{array}
\right. , \nonumber \\
\label{eq:app.repliconmode}
\end{eqnarray}
for two specific replicas $\theta$ and $\mu$. 
In the limit $n\to 0$, perturbations keep symmetry of the eigenvectors $\vec{\mu}_1$ and $\vec{\mu}_2$ across the replicas. 
Therefore, $\vec{\mu}_1$ and $\vec{\mu}_2$ are irrelevant to replica symmetry breaking (RSB) but only determines the stability within the RS ansatz. 
Hence, the third eigenvector $\vec{\mu}_3$, which is called the replicon mode, causes RSB. 
The eigenvalue equation $\hat{G}\vec{\mu}_3=\lambda_3\vec{\mu}_3$ with respect to (\ref{eq:app.repliconmode}) splits into $T\vec{\mu}_3=\vec{0}$ and $[U+(K-1)V]\vec{\mu}_3'=\lambda_3\vec{\mu}_3'$, where $\vec{\mu}_3 = {}^t(\vec{0}, \vec{\mu}_3')$. 
Therefore, the eigenproblem of $\hat{G}$ is equivalent to that of $U+(K-1)V$. 
\par
Let us calculate the elements of $U$ and $V$. 
The second derivative ${\cal M}$ by $q_l^{ab}$ related to the $U^{ab,cd}, V^{ab,cd}$ is 
\begin{eqnarray}
\frac{\partial^2 {\cal M}}{\partial q_l^{ab} \partial q_{l'}^{cd}} 
&=& R^{-1}<v_l^a v_l^b v_{l'}^c v_{l'}^d>_{u,v} \nonumber \\
& & - R^{-1}<v_l^a v_l^b>_{u,v} <v_{l'}^c v_{l'}^d>_{u,v}, 
\label{eq:app.PQRP'Q'R'}
\end{eqnarray}
where 
\begin{widetext}
\begin{equation}
<g(\{ v_l^a \})>_{u,v}=\frac
{\displaystyle \biggl< \int \biggl( \prod_l d\vec{u}_l d\vec{v}_l e^{-\frac 12 {}^t\vec{v}_l Q_l \vec{v}_l + i\vec{v}_l\cdot\vec{u}_l} \biggr) g(\{ v_l^a \}) \prod_a \biggl\{ e^{-\beta^a} + (1-e^{-\beta^a}) \Theta ( y,\{u_l^a\}) \biggr\} \biggr>_y}
{\displaystyle \biggl< \int \biggl( \prod_l d\vec{u}_l d\vec{v}_l e^{-\frac 12 {}^t\vec{v}_l Q_l \vec{v}_l + i\vec{v}_l\cdot\vec{u}_l} \biggr) \prod_a \biggl\{ e^{-\beta^a} + (1-e^{-\beta^a}) \Theta ( y,\{u_l^a\}) \biggr\} \biggr>_y}, 
\end{equation}
\end{widetext}
for any function $g(\{ v_l^a \})$. 
The second derivative ${\cal M}$ by $\hat{q}_l^{ab}$ related to the $\hat{U}^{ab,cd}, \hat{V}^{ab,cd}$ is 
\begin{eqnarray}
\frac{\partial^2 {\cal M}}{\partial \hat{q}_l^{ab} \partial \hat{q}_{l'}^{cd}}
&=& 
\left\{
\begin{array}{ll}
K^{-1}<s^a s^b s^c s^d>_s & \\
\quad - K^{-1}<s^a s^b>_s <s^c s^d>_s, & \; (l=l'), \\
0, & \; (l\ne l'),
\end{array}
\right. \nonumber \\
& & 
\label{eq:app.P^Q^R^P^'Q^'R^'}
\end{eqnarray}
where 
\begin{equation}
<g(\{ s^a \})>_s=\frac
{\displaystyle \int Dz \mathop{\rm Tr}_{\{ s^a\}} g(\{ s^a \}) \exp \biggl( \sqrt{\hat{q}}\,z \sum_a s^a \biggr)}
{\displaystyle \int Dz \mathop{\rm Tr}_{\{ s^a\}} \exp \biggl( \sqrt{\hat{q}}\,z \sum_a s^a \biggr)}, 
\end{equation}
for any function $g(\{ s^a \})$. 
The second derivative ${\cal M}$ by $q_l^{ab},\hat{q}_l^{ab}$ related to the $\tilde{U}^{ab,cd}, \tilde{V}^{ab,cd}$ is 
\begin{equation}
\frac{\partial^2 {\cal M}}{\partial q_l^{ab} \partial \hat{q}_{l'}^{cd}}
=
\left\{
\begin{array}{ll}
K^{-1}, & \; (l=l', a=c, b=d), \\
0, & \; ({\rm otherwise}).
\end{array}
\right.
\label{eq:app.PtQtRtPt'Qt'Rt'}
\end{equation}
Using Gardner's method \cite{Gardner1988}, we find that the RS stability criterion is 
\begin{equation}
K \gamma <1,
\label{eq:app.AT}
\end{equation}
where 
\begin{eqnarray}
\gamma   &\equiv& \gamma_0 + (K-1) \gamma_1, \nonumber \\
\gamma_0 &\equiv& P-2Q+R, \nonumber \\
\gamma_1 &\equiv& P'-2Q'+R', \nonumber \\
P  &\equiv& U^{ab,ab}, \nonumber \\
Q  &\equiv& U^{ab,ac} \quad (b\ne c), \nonumber \\
R  &\equiv& U^{ab,cd} \quad (a\ne c,b\ne d), \nonumber \\
P' &\equiv& V^{ab,ab}, \nonumber \\
Q' &\equiv& V^{ab,ac} \quad (b\ne c), \nonumber \\
R' &\equiv& V^{ab,cd} \quad (a\ne c,b\ne d). 
\end{eqnarray}
The line $K \gamma =1$ is called the AT line. 
Setting $K=0$, on the other hand, the matrix $U+(K-1)V$ is equal to $U-V$. 
When $K=0$, inequality $K\gamma =0<1$ of (\ref{eq:app.AT}) always holds. 
Therefore, permutation symmetry breaking (PSB) does not occur in this system.

\subsection{For lossy compression using a parity tree with $K = 2$ hidden units} 
\label{appendix.ATline.K=2PT}

Let us consider the RS stability of lossy compression using a parity tree with $K=2$ hidden units. 
Here, $\Theta ( y,\{u_l^a\})$ is given by $\Theta ( y,\{u_l^a\}) = \Theta ( y\prod_l \sgn (u_l^a))$, therefore solutions to the saddle-point equations are 
\begin{equation}
q=\hat{q}=0, \; D=\frac{e^{-\beta}}{1+e^{-\beta}}. 
\label{eq:app.K>=2PTspe}
\end{equation}
Substituting (\ref{eq:app.K>=2PTspe}) into (\ref{eq:app.PQRP'Q'R'}) and (\ref{eq:app.P^Q^R^P^'Q^'R^'}), we obtain 
\begin{eqnarray}
P'&=& R^{-1}\frac 4{\pi^2}(1-2D)^2, \nonumber \\
P &=& Q=R=Q'=R'=0. 
\end{eqnarray}
Therefore, using (\ref{eq:app.AT}), the RS stability can be obtained as 
\begin{equation}
R > \frac 8{\pi^2}(1-2D)^2 \equiv R_{AT}(D). 
\end{equation}
This proves (\ref{eq:ATline}).

\subsection{For lossy compression using a parity tree with $K \ge 3$ hidden units} 
\label{appendix.ATline.K>=3PT}

Next, let us consider the RS stability of lossy compression using a parity tree with $K \ge 3$ hidden units. 
Here, the solutions to the saddle-point equations are $q=\hat{q}=0, D=e^{-\beta}/(1+e^{-\beta})$ as well as for $K=2$. 
Substituting (\ref{eq:app.K>=2PTspe}) into (\ref{eq:app.PQRP'Q'R'}) and (\ref{eq:app.P^Q^R^P^'Q^'R^'}), we obtain 
\begin{equation}
P=Q=R=P'=Q'=R'=0. 
\end{equation}
Since the inequality $K \gamma =0<1$ of (\ref{eq:app.AT}) always holds, 
the RS solution does not exhibit the AT instability throughout the achievable region for $K\ge 3$.

\section{A lower bound to the rate distortion property of lossy compression using a parity tree} 
\label{appendix.lower_bound}

In order to derive a lower bound to the rate distortion property, an upper bound to the entropy is necessary. 
Using Jensen's inequality, an upper bound to the entropy $S^{\; upper}(D,R)$ is given by 
\begin{align}
S(D,R) 
&= \frac 1N <\ln {\cal N}(D,R)>_{\vec{y},\vec{x}} \nonumber \\
&\le \frac 1N \ln <{\cal N}(D,R)>_{\vec{y},\vec{x}} \nonumber \\
&\equiv S^{\; upper}(D,R). 
\end{align}
After a simple calculation, we obtain the upper bound to the entropy of lossy compression using a parity tree $S_{PT}^{\; upper}(D,R)$ as 
\begin{align}
S_{PT}^{\; upper}(D,R) 
=& \ln 2 + \mathop{\rm extr}_{\beta} \biggl( R^{-1} \ln \frac{1+e^{-\beta}}2 + \beta R^{-1}D \biggr) \nonumber \\
=& -R^{-1}\ln 2+ \ln 2 -R^{-1}D\ln D \nonumber \\
 & -R^{-1}(1-D)\ln (1-D). 
\end{align}
Note that this annealed entropy $S_{PT}^{\; anneal}(D,R)$ is not depend on the number of hidden units $K$. 
Solving $S_{PT}^{\; anneal}(D,R)=0$ with respect to $R$, we obtain 
\begin{equation}
R = 1-h_2(D). 
\end{equation}
This shows that the rate distortion function for uniformly unbiased binary sources (\ref{eq:RDF}) can be also derived 
as a lower bound to the rate distortion property of compression using a parity tree. 
\par
We next discuss a upper bound to the rate distortion property. 
In order to derive a upper bound to the rate distortion property, we need an lower bound to the entropy. 
Using Jensen's inequality, an upper bound to the entropy $S^{\; upper}(D,R)$ is represented by 
\begin{align}
S(D,R) 
&= \frac 1N < \ln {\cal N}(D,R)>_{\vec{y},\vec{x}} \nonumber \\
&= \frac 1N \biggl< - \ln \frac 1{{\cal N}(D,R)} \biggr>_{\vec{y},\vec{x}} \nonumber \\
&\ge - \frac 1N \ln \biggl< \frac 1{{\cal N}(D,R)} \biggr>_{\vec{y},\vec{x}} \nonumber \\
&\equiv S^{\; lower}(D,R). 
\end{align}
This inequality can be also obtained by an annealed information entropy as follows. 
According to Opper's discussion \cite{Opper1995}, we first define a function that characterizes a version space as follows: 
\begin{align}
\rho(\vec{s}) \equiv \frac{\delta \biggl( MD ; d(\vec{y},\hat{\vec{y}}(\vec{s})) \biggr)}{\displaystyle{\mathop{\rm Tr}_{\vec{s}}\delta \biggl( MD ; d(\vec{y},\hat{\vec{y}}(\vec{s})) \biggr)}}. 
\end{align}
Since this function $\rho(\vec{s})$ is non-negative and normalized to $\mathop{\rm Tr}_{\vec{s}} \rho(\vec{s})=1$, it defines a probability with respect to $\vec{s}$. 
Therefore we obtain the information entropy per bit ${\cal H}(D,R)$ as 
\begin{align}
{\cal H}(D,R)
\equiv& \frac 1N \biggl< \mathop{\rm Tr}_{\vec{s}} \rho(\vec{s}) \ln \frac 1{\rho(\vec{s})} \biggr>_{\vec{y},\vec{x}} \nonumber \\
=     & \frac 1N \biggl< \ln \frac 1{\rho(\vec{s})} \biggr>_{\vec{s},\vec{y},\vec{x}} \nonumber \\
\ge   & - \frac 1N \ln < \rho(\vec{s}) >_{\vec{s},\vec{y},\vec{x}} \nonumber \\
=     & - \frac 1N \ln < \mathop{\rm Tr}_{\vec{s}} \rho(\vec{s})^2 >_{\vec{y},\vec{x}} \nonumber \\
=     & - \frac 1N \ln \biggl< \frac 1{{\cal N}(D,R)} \biggr>_{\vec{y},\vec{x}}, 
\end{align}
where $< g(\vec{s}) >_{\vec{s}} = \mathop{\rm Tr}_{\vec{s}} \rho(\vec{s}) g(\vec{s})$. 
Using the identity 
\begin{align}
\rho(\vec{s}) \ln \frac 1{\rho(\vec{s})}=
\left\{
\begin{array}{l}
0, \quad {\rm if} \; \delta ( MD ; d(\vec{y},\hat{\vec{y}}(\vec{s})) )=0, \\
{\cal N}(D,R)^{-1} \ln {\cal N}(D,R), \; {\rm otherwise},
\end{array}
\right.
\end{align}
we can easily confirm ${\cal H}(D,R)=S(D,R)$. 
\par
However, it is difficult to evaluate the lower bound $S^{\; lower}(D,R)$ directly because $< {{\cal N}(D,R)}^{-1} >_{\vec{y},\vec{x}} \ge < {{\cal N}(D,R)} >_{\vec{y},\vec{x}}^{-1}$. 
This difficulty is caused by a limitation of the version space due to the distortion. 
This limitation complicates the probability $\rho(\vec{s})$.


\end{document}